\documentclass[twocolumn,showpacs,preprintnumbers,amsmath,amssymb,superscriptaddress,bibnotes,floatfix]{revtex4-1}

\usepackage{graphicx}% Include figure files
\usepackage{dcolumn}% Align table columns on decimal point
\usepackage{bm}% bold math
\begin{document}	
\preprint{APS/123-QED}
\title{Temporal and structural heterogeneities emerging in adaptive temporal networks}

\author{Takaaki Aoki}
\email{takaaki.aoki.work@gmail.com}
\affiliation{Faculty of Education, Kagawa University, Takamatsu, Japan}
\author{Luis E. C. Rocha}
\affiliation{Department of Mathematics, Universit\'e de Namur, Namur, Belgium}
\affiliation{Department of Public Health Sciences, Karolinska Institutet, Stockholm, Sweden}
\author{Thilo Gross}
\affiliation{Department of Engineering Mathematics, University of Bristol, Bristol, UK}

\date{\today}

\begin{abstract}
We introduce a model of adaptive temporal networks whose evolution is regulated by an interplay between node activity and dynamic exchange of information through links. We study the model by using a master equation approach. Starting from a homogeneous initial configuration, we show that temporal and structural heterogeneities, characteristic of real-world networks, spontaneously emerge. This theoretically tractable model thus contributes to the understanding of the dynamics of human activity and interaction networks.
\end{abstract}
\pacs{05.65.+b, 89.75.Fb, 89.75.Hc}
\maketitle

%%%%%%%%%%%%%%%%%%%%%%%%%%%
% INTRODUCTION
%%%%%%%%%%%%%%%%%%%%%%%%%%%

Human social behaviour depends both on intrinsic properties of the individuals and on the interactions between them. In daily life, interactions between people create contact patterns that can be mathematically represented by networks, i.e.\ a set of nodes, corresponding to the people, connected by links, representing the contacts between the respective individuals \cite{NewmanBook2010}. In network terminology, the number of contacts of a given person is called the degree $k$ of a node and thus the degree distribution $p_k$ is the probability distribution that a randomly chosen node has degree $k$. A central observation is that most real-life networks have a high level of heterogeneity in the number of contacts per node and the empirical degree distributions are typically approximated by power-laws $p_k\sim k^{-\gamma}$ \cite{Albert:2002p869,Newman:2003p839,UganderFacebook}. 

A second observation is that human contact networks are not static. For instance in studies of email contact networks, users that are hubs of the network in one time window may be unremarkable or even isolated in the next time window \cite{Hill2010}.
This may reflect the fact that nodes alternate between an active, contact-seeking state, and an inactive states.
The temporal heterogeneity can then be quantified in terms of the inter-event intervals (IETs) between node activations,
which reveals burstiness of human behavior \cite{Barabasi:2005aa,Alexei2006PRE}.

While several models have been proposed to explain the heterogeneity in the degree distribution \cite{Barabasi:1999p178,Caldarelli2002PRL,Papadopoulos:2012aa,Muchnik2013}, fewer studies focused on modeling the burstiness of the temporal activity \cite{Barabasi:2005aa,Alexei2006PRE,Vajna2013}. Barab\'asi proposed a priority-based model in which nodes first execute the high-priority tasks, i.e.\ these tasks are executed within a short time, while low-priority tasks have to wait longer times before leaving the queue. Other models use inhomogeneous Poisson processes on each node modulated by (daily and weekly) cycles of human activity \cite{Malmgren25112008,HidalgoR2006877}. Combined, these processes generate the patterns of burstiness that are comparable to real world data.

While previous models thus describe network heterogeneity and burstiness as different phenomena, they are connected in the real world: Network structure 
is a result of the activity of the network nodes, while changes in activity are likely to be triggered by neighbours in the network. The system is thus modelled as an adaptive network \cite{Sayama20131645, Gross:2008p274}, where dynamics of nodes is  affected by network structure, while the evolution of the structure is dependent on the state of the nodes. 
 
A number of models have been proposed \cite{Holme201297,Holme2015},
in which the network links are temporal and change according to state of nodes \cite{Zha02011,Starnini2013}.
In particular, an idea that temporal links are generated according to activity parameter of nodes that is obtained by empirical data was independently presented in Refs. \cite{Rocha:2011bq,Perra:2012fi}.
This modeling framework is extended to incorporate the memory effect of the past contacts \cite{Karsai:2014aa,Vestergaard2014}, which can be also regarded as a type of adaptive networks.

\begin{figure}[h]
\begin{center}
\includegraphics[]{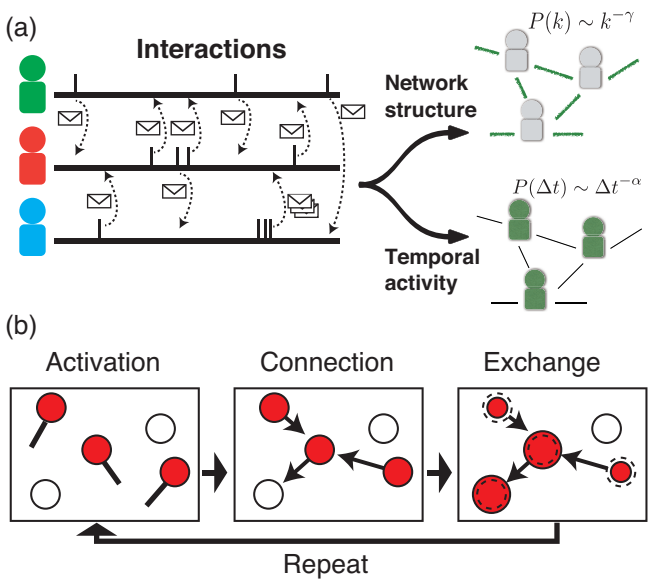}
\caption{(Color online) (a) Question : How do social interactions induce structural and temporal heterogeneities among people?
 (b) Illustration of the interaction-regulated stochastic contact model. Within one time step, (i) nodes become activate, (ii) make random connections, (iii) exchange resources, and finally (iv) break down the links. 
 %(c) Two types of inter-event intervals (IETs): single-node and population.
 }
\label{fig:Scheme}
\end{center}
\end{figure}

In this paper, using the framework of  Refs. \cite{Rocha:2011bq, Perra:2012fi},  
we consider an  interplay between individual activity and network structure: the human activity 
is dynamically determined by the state of the node, and simultaneously the state is updated by the  contacts between the nodes, as illustrated in Fig.~\ref{fig:Scheme}(a). 
Our toy model is motivated by interactions in online environments, particularly web-forums \cite{RochaPNAS2010} and dating sites \cite{Holme2003EPL}, where there is an interplay between exchange of information (i.e. messages) and activity (i.e. posting) of the members, such that past interactions trigger new events.
We analyze the master equation of the model using generating functions. We show that, starting from homogeneous initial conditions, structural heterogeneity and temporal burstiness can emerge spontaneously.

%%%%%%%%%%%%%%%%%%%%%%%%%%%
% MODEL DESCRIPTION
%%%%%%%%%%%%%%%%%%%%%%%%%%%

We consider a population of $N$ nodes, where each node $i$ has an intrinsic variable $x_i$, which we interpret as an abstract resource representing the node's willingness to engage with others in the network. Initially every node is assigned the same resource quantity,  i.e., $x_i(0) =1$.  
The system then evolves due to dynamical updates which comprise three steps: (i) activation of nodes, (ii) formation of pairs, and (iii) exchange of resources (Fig.~\ref{fig:Scheme}(b)).

In the activation step (i) $N_A$ nodes are set to the active state, while all others are set to the inactive state. The active nodes are chosen initially using a linear 
preferential rule, such that the probability that node $i$ becomes active scales with $x_i$.
In the pair formation step (ii) every active node picks a partner. 
With probability $\kappa$ this partner is chosen randomly from all nodes in the system. 
With the complementary probability $1-\kappa$ the partner is chosen randomly from the set of active nodes. 
This also reduces to previously studied rules in the limit $\kappa=0$ \cite{Rocha:2011bq}, where partners are only picked among the active nodes, and in the limit $\kappa = 1$ \cite{Perra:2012fi}, where partners are picked among the whole network. 
Finally, in the exchange step nodes transfer an amount of resource to their partners such that 
\begin{align}
x_i(t+1) - x_i(t) =   D \left[ \sum_{j} a_{ij}(t)  - \sum_{j} a_{ji}(t) \right],
\end{align}
where $D$ is the amount of resource transferred in the interaction, and
$a_{ij}(t)=1$ if $i$ has picked $j$ as its partner, and 0 otherwise.

%% I added the following texts to explain the realistic interpretation of the model
Throughout above procedure, this model mimics social interactions in online, web-forums and SNSs.
Users become active according to their willingness to engage with others denoted by $x_i$, and either comment on someone's else threads (random connections over the network) or respond to posts of other active (connections to active nodes) members. This will increase the willingness of receivers who get the comments, whereas it will satisfy the demand of the senders (resource exchange).

%%%%%%%%%%%%%%%%%%%%%%%%%%%
% SHOW RICH BEHAVIOR OF THE MODEL NUMMERICALLY
%%%%%%%%%%%%%%%%%%%%%%%%%%%
Firstly, we demonstrate the self-organization of the system.
We consider the model from a network perspective. Node $i$ picking a partner $j$ constitutes the creation of a directed link from $i$ to $j$ at every steps. Thus the matrix ${\rm \bf a(t)}$ is interpreted as a directed adjacency matrix.  As a result of the temporal contacts between nodes, the initial identical distribution of resources on nodes become heterogeneous over time. This leads to the emergence of a dynamic network with structural heterogeneity and burstiness of node activations, as shown in Fig. \ref{fig:Diagram} (a) 
(see movie for a dynamic behavior of the model  \footnote{See Supplemental Material at [URL will be inserted by publisher] for the movie}).

By tuning the parameter $\kappa$, this model is able to reproduce different structural and temporal patterns (Fig. \ref{fig:Diagram}).
In the figure, we show the degree distribution of the aggregated network which is formed by all links collected during $T_s = 10^4$ time steps,
and the IET distribution of a single node during $5 \times 10^7$ steps.
If $\kappa = 0$, a small group of active nodes emerges. Nodes outside this group are left without resources.
In this situation, the IET distribution of the active nodes exhibits a Poisson-like dynamics, i.e.~exponential inter-event times (Fig. \ref{fig:Diagram}(c)).
By contrast, sufficiently large $\kappa$ leads to a homogeneous distribution of resources, which generates a Gaussian-like in-degree distribution (Fig. \ref{fig:Diagram}(f)) and an exponential (single-node) IET distribution (Fig. \ref{fig:Diagram}(g)), the later a result of the quasi-homogenous Poisson process.
In the intermediate case, where active nodes mainly link to other active individuals but also occasionally inactive ones, we observe the emergence of highly heterogeneous structure (Fig. \ref{fig:Diagram}(d)) and temporal patterns (Fig. \ref{fig:Diagram}(e)). 
Owing to the period for aggregation $T_s (= 10^4)$, the network has an upper limit of the degree and its distribution has an exponential cutoff.
In this situation,  the distributions of the weights and the IETs at a link level are also very heterogenous, and can be fitted by a power-law (data not shown).

\begin{figure*}[tbp]
\begin{center}
\includegraphics[]{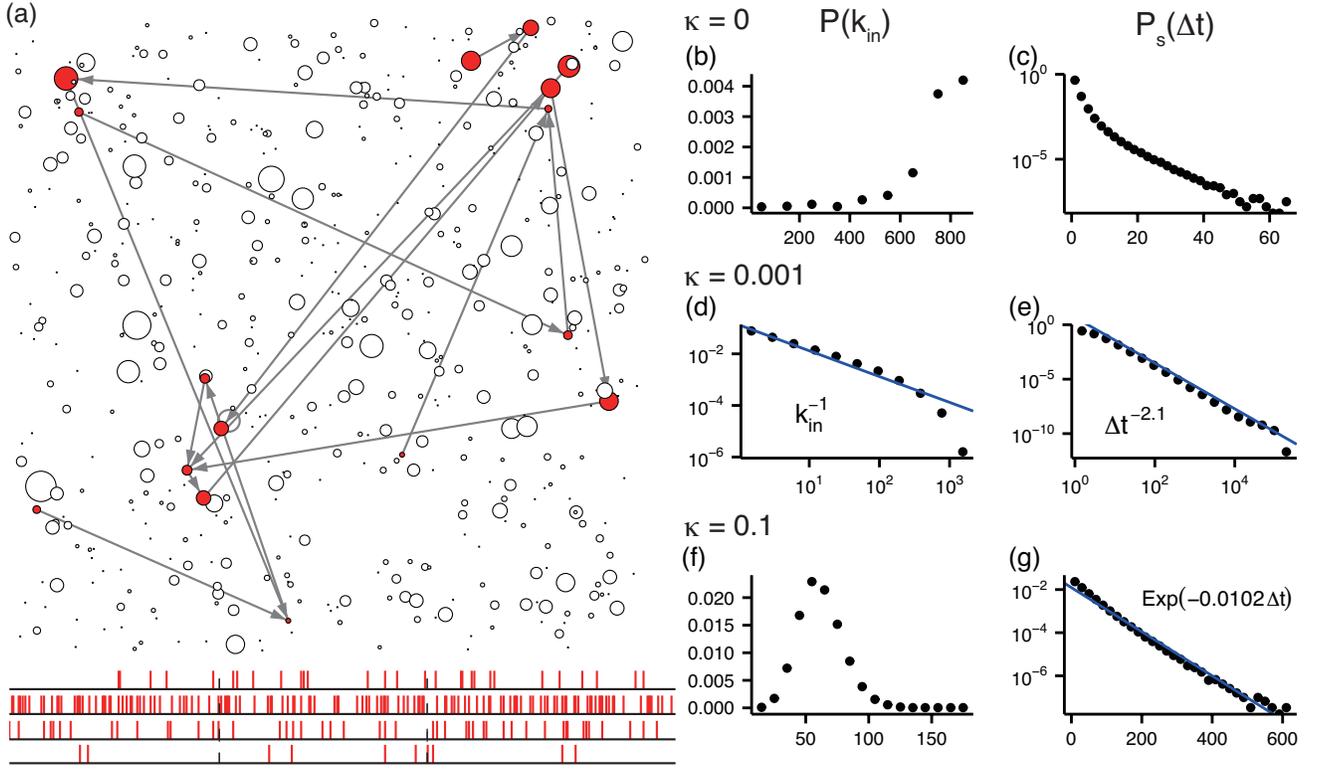}
\caption{(Color online)
(a) A snapshot of  dynamic behavior of the model. See movie for details in Supplemental Material.
(b-g) The degree distribution $P(k)$ and single-node IETs distribution $P_s(\Delta t)$ for (b,c) $\kappa = 0$ ($P_s(\Delta t)$ is plotted in semi-log graph), (d,e) $\kappa = 0.001$ (both graphs are log-log), and (f,g) $\kappa = 0.1$ ($P_s(\Delta t)$ is semi-log), with $N_A = 1024$, $N = 2^{15}$, and $D=0.01$. 
}
\label{fig:Diagram}
\end{center}
\end{figure*}

%%%%%%%%%%%%%%%%%%%%%%%%%%%
% ANALYSIS / MASTER EQUATION
%%%%%%%%%%%%%%%%%%%%%%%%%%%

To make theoretical progress we construct a master equation for the resource 
dynamics. We define $u_n(t)$ as the density of nodes $i$ at resource level $x_i = nD$.
Setting the total resource in the system to $\sum n D u_n=1$ and assuming large $N$ and $N_A$, the probability that a node at resource level $n$ becomes active is $a_n=(N_A/N)(nD/\sum_n n D u_n) = nDN_A/N$. Therefore, the proportion of nodes that are both 
active and at resource level $n$ is $a_n u_n = N_A n D u_n / N$. 

Since the total number of active nodes is $N_A$, $N_A$ links are formed in every time step.
Of these $M1:=N_A \kappa$ links point to random targets chosen among the whole population, whereas $M2:=N_A (1-\kappa)$ point to targets chosen only among the active nodes.
From the perspective of a single node the placement of a link can be seen as a statistical trial that is successful if that specific node is chosen as the target of the link. 
Every node, irrespective of state, receives a link with probability $\rho_1=1/N$ in each
of the $M_1$ trials where the targets are random nodes. In addition, active nodes receive 
a link with probability $\rho_2 = 1/ N_A$ in each of the $M_2$ trials where the targets are random active nodes. 

Each node then gains $D$ units of resource for every incoming link, while active nodes 
lose $D$ units of resource via their outgoing link. Using the Binomial distribution $B(m, \rho, M)$ of $m$ success in $M$ trials, then leads to the master equation
\begin{align}
\frac{d u_n}{d t} &= A(-1) a_{n+1} u_{n+1}     \notag \\ 
 &   -  C_1 a_n u_n  - C_2 \bar{a}_n u_n  \notag \\
 & +  \sum_{m = 1}^{N_A}  a_{n - m} u_{n - m}A(m)  \notag \\
 &   +\sum_{m = 1}^{N_A \kappa} \bar{a}_{n-m} u_{n - m} B(m, \rho_1, M_1) , \label{eq:master}
\end{align}
where we now treat time continuously and $\bar{a}_n (= 1 -a_n)$ is the inactive fraction of $u_n(t)$, $A(m) = \sum_{m_1 + m_2 = m + 1} B(m_1, \rho_1, M_1) B(m_2, \rho_2, M_2)$, and $C_1 = 1 - A(0)$, $C_2 = \sum_{m=1}^{N_A \kappa} B(m_1, \rho_1, N_A \kappa)$.

%%%%%%%%%%%%%%%%%%%%%%%%%%%
% ANALYSIS / GENERATING FUNCTION
%%%%%%%%%%%%%%%%%%%%%%%%%%%

For the analysis of the master equation it is useful to write a generating function
$Q(t,x) = \sum_n u_n(t) x^n$ \cite{Generatingfunctionology} \footnote{See the supplementary material for details on the generating function}. This function encodes the $u_n$ in a continuous function 
by interpreting them as Taylor coefficients in an abstract variable $x$, which does not have 
a physical meaning. Multiplying equation \eqref{eq:master} by $x^n$ and summing over $n \ge 0$
we obtain
\begin{align}
\frac{\partial Q}{\partial t} & =  \frac{N_A D}{N} \frac{\partial Q}{\partial x} \left[  A(-1) +  (- C_1  + C_2) x  + \sum_{m=1}^{N_A-1} A(m) x^{m+1}  \right. \notag \\
& \left.    - \sum_{m=1}^{N_A\kappa} B(m, \rho_1, M_1) x^{m+1} \right]  \notag \\
 &+ Q \left[-C_2  + \sum_{m=1}^{N_A\kappa} B(m, \rho_1, M_1) x^m \right].
\end{align}
In the limit $N, N_A \to \infty$, we approximate the Binomial distributions reduce to Poisson distributions with finite rates,  $\lambda_1 (= \rho_1 M_1 = \frac{N_A}{N} \kappa )$ and $\lambda_2 (= \rho_2 M_2 = 1 - \kappa)$, respectively. We obtain
\begin{align}
\frac{\partial Q}{\partial t}  = \frac{N_AD}{N}  Y(x) \frac{\partial Q}{\partial x}  + Z(x) Q,  \label{eq:pde}
\end{align}
where $Y(x) =   \exp \left( (\lambda_1+\lambda_2)(x-1)\right)   - x \exp \left(\lambda_1(x-1)\right)$ and $Z(x) = -1 + \exp\left(\lambda_1 (x -1) \right)$.
Thus the generating approach has converted the large system of ordinary differential 
equations to a single partial differential equation.  

\begin{figure}[tb]
\begin{center}
\includegraphics[]{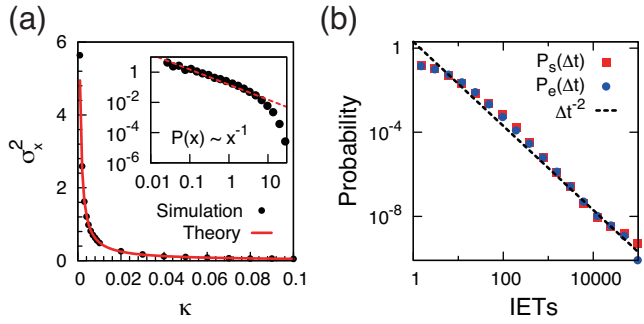}
\caption{(Color online) (a) The  variance $\sigma^2_x$ of the stationary resource distribution given by Eq. \eqref{eq:variance} (red line) and for the simulation of the agent-based model (black circles). We use $D = 0.01$, $N_A = 1000$,  and $N=2^{15}$. 
The inset shows the resource distribution for $\kappa = 0.001$.
(b) The distributions of population and single-node IETs,  respectively $P_e(\Delta t)$ and $P_s(\Delta t)$. The master equation analysis predicts  $P_e (\Delta t) \propto \Delta t^{-2}$.}
\label{fig:Gen}
\end{center}
\end{figure}

When considered in the stationary state, Eq.~(\ref{eq:pde}) relates $Q$ to its own first derivative $Q'$. For any probability distribution the corresponding generating function 
must obey $Q(1)=1$. We can therefore find $Q'(1)$ and by differentiating $Q''(1)$.    
These quantities are of interest because $Q'(1)=\mu$ is the mean of $u_n$ 
and $Q''(1)$ is closely related to the variance $\sigma^2=Q''(1)+Q'(1)-{Q'(1)}^2$ of $u_n$. 
From this we can obtain the mean $\mu_x$ and variance $\sigma_x^2$ of the resource distribution 
\begin{align}
\mu_x &= 1, \\
\sigma^2_x  & =\frac{D}{2  \kappa} \left[ 1  -2  \kappa + \left(1 - \frac{N_A}{N}\right) \kappa^2 \right] + D . \label{eq:variance}
\end{align}
In the case where targets are almost always chosen among the active nodes ($\kappa \sim 0), $ we find 
the resource amounts among nodes will be extremely heterogeneous (Fig.~\ref{fig:Gen}(a)). 
This result is also found in agent-based simulations as shown in the inset of Fig.~\ref{fig:Gen}(a), 
in which the resource distribution is well fitted by a power-law $x^{-1}$ in some ranges.
We further confirmed that the same scaling behaviour also appears in 
a continuous-time version of the model (not shown). These results are interesting since 
the power law exponent $\gamma=1$ differs from  $2<\gamma<3$ that is reported to most scale-free networks in empirical studies \cite{NewmanBook2010, Boccaletti2006175}.
At present we do not have an explanation for this exponent and cannot strictly exclude
that the $1/k$ behaviour is an extremely long transient.  

The master equation looses information on the temporal activity of the nodes. Instead of single-node IETs $P_s(\Delta t)$, we thus consider the population IETs $P_e(\Delta t)$.%(Fig.~\ref{fig:Scheme}(c)). 
If the amount of resources is fixed, the activation of $u_n^*$ can be described by a Poisson process with a fixed rate $a_n$.
In general, the population IETs of homogeneous Poisson nodes $\{p_i\}$ $(i=1,2,\cdots,N)$ is given by \cite{HidalgoR2006877}
\begin{align}
P_e(\Delta t) = \sum_i p_i e^{-p \Delta t}  = \int f(p) p e^{-p \Delta t} dp, \notag
\end{align}
where $f(p)$ is the rate distribution in the limit of $N \to \infty$. In this equation, only the first time-interval of the node's activations is collected for each node \cite{HidalgoR2006877}. The second, third and succeeding intervals should be taken into account for population IETs during a given observation period. The higher the rate of a node, the more likely the IET data is collected from this node. The probability of the IET should be multiplied by the rate $p$, and then the equation is rewritten as
\begin{align}
P_e(\Delta t) = \int f(p) p^2 e^{-p \Delta t} dp \propto \Delta t^{-2}. \label{eq:pIETs}
\end{align}
In the last part, we considered  the case of $f(p) \propto p^{-1}$, because $u_n^*$ follows the power-law with exponent $-1$. Figure \ref{fig:Gen}(b) shows the distributions of the population IETs with the observation period $T_o = 10^5$ and of the single-node IETs of a uniformly sampled node, obtained by the direct simulation of the model with the same parameters as in Fig. \ref{fig:Gen}(a). We find that both IETs, $P_e(\Delta t)$ and $P_s(\Delta t)$, are close to the power-law predicted theoretically.

%%%%%%%%%%%%%%%%%%%%%%%%%%%
% EXTENTION : NONLINEAR VER.
%%%%%%%%%%%%%%%%%%%%%%%%%%%
\begin{figure}[tbp]
\begin{center}
\includegraphics[]{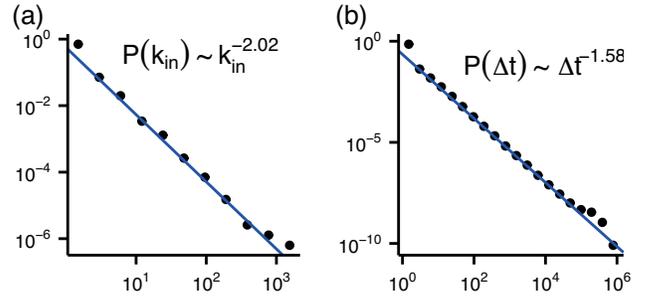}
\caption{(Color online) Effect of nonlinear  activation probability.
(a) The degree distribution $P(k)$. (b) single-node IETs distribution $P_s(\Delta t)$ (log-log graphs).
We use $\alpha$ = 2, $\kappa$ = 0.0001, $D$ =1, $N_A$ = 512, and $N$=8192.}
\label{fig:Nonlinear}
\end{center}
\end{figure}

The resource amount of a node $x_i$ determines its activation probability $a_i$, i.e.  $a_i \propto x_i$.  A natural extension is to consider nonlinear preferential connections ($a_i \propto x_i^\alpha$) \cite{RochaPNAS2010}. In some situations there is an advantage of large scale, in which accumulated resource enhances super-linearly the activity of a person ($\alpha > 1$). On the other hand, there may be saturation of the effect of the resources, causing sub-linear activity ($\alpha < 1$).
Figure 4 shows that the nonlinear preferential rule also produces the power-law distributions of node degree and IETs, but with different exponents. The power-law exponent is about 2 for the degree distribution and about 1.5 for IETs distribution if $\alpha=2$.
Our findings are consistent with empirical results of online communication in a web-forum \cite{RochaPNAS2010} and with correspondence patterns (waiting times) of famous scientists \cite{Alexei2006PRE}, reinforcing the intuition that some classes of communication processes are regulated by feedback mechanisms between information available and human activity.
 
%%%%%%%%%%%%%%%%%%%%%%%%%%%
% DISCUSSIONS
%%%%%%%%%%%%%%%%%%%%%%%%%%%

In this paper, we proposed a model of contact networks where the human dynamics are adaptively regulated by past interaction and exchange of resources. 
We analyzed the master equation of the model and found structural and temporal heterogeneities which are remarkable properties observed in many real-world systems.
This revealed that, these two types of heterogeneity can be observed in rich-club-like systems where the active individuals typically communicate with other active individuals, but occasionally connect to anyone in the network.

Despite the relative simplicity, the model provides a potential mechanistic understanding of the online communication dynamics, pointing to a number of unsolved questions, including the exact nature of the observed transition.
Moreover, there are other important structural and temporal properties that our model can not explain, such as community structure \cite{Clauset:2004dz,Newman:2006iq,Fortunato:2010iw}.
We therefore hope that it will stimulate future work, leading to deeper insights in the behavioral patterns of humans.

%%%%%%%%%%%%%%%%%%%%%%%%%%%
% ACKNOWLEDGEMENTS
%%%%%%%%%%%%%%%%%%%%%%%%%%%

We thank T. Takaguchi, N. Perra, N. Masuda, and S. Shinomoto for fruitful discussions. This work was supported by JSPS KAKENHI Grants No. 24740266 and No. 26520206 and by Bilateral Joint Research Projects between JSPS and F.R.S.-FNRS. LECR is a FNRS Charg\'e de recherche and thanks Hierta-Retzius Foundation for financial support.


\begin{thebibliography}{33}%
\makeatletter
\providecommand \@ifxundefined [1]{%
 \@ifx{#1\undefined}
}%
\providecommand \@ifnum [1]{%
 \ifnum #1\expandafter \@firstoftwo
 \else \expandafter \@secondoftwo
 \fi
}%
\providecommand \@ifx [1]{%
 \ifx #1\expandafter \@firstoftwo
 \else \expandafter \@secondoftwo
 \fi
}%
\providecommand \natexlab [1]{#1}%
\providecommand \enquote  [1]{``#1''}%
\providecommand \bibnamefont  [1]{#1}%
\providecommand \bibfnamefont [1]{#1}%
\providecommand \citenamefont [1]{#1}%
\providecommand \href@noop [0]{\@secondoftwo}%
\providecommand \href [0]{\begingroup \@sanitize@url \@href}%
\providecommand \@href[1]{\@@startlink{#1}\@@href}%
\providecommand \@@href[1]{\endgroup#1\@@endlink}%
\providecommand \@sanitize@url [0]{\catcode `\\12\catcode `\$12\catcode
  `\&12\catcode `\#12\catcode `\^12\catcode `\_12\catcode `\%12\relax}%
\providecommand \@@startlink[1]{}%
\providecommand \@@endlink[0]{}%
\providecommand \url  [0]{\begingroup\@sanitize@url \@url }%
\providecommand \@url [1]{\endgroup\@href {#1}{\urlprefix }}%
\providecommand \urlprefix  [0]{URL }%
\providecommand \Eprint [0]{\href }%
\providecommand \doibase [0]{http://dx.doi.org/}%
\providecommand \selectlanguage [0]{\@gobble}%
\providecommand \bibinfo  [0]{\@secondoftwo}%
\providecommand \bibfield  [0]{\@secondoftwo}%
\providecommand \translation [1]{[#1]}%
\providecommand \BibitemOpen [0]{}%
\providecommand \bibitemStop [0]{}%
\providecommand \bibitemNoStop [0]{.\EOS\space}%
\providecommand \EOS [0]{\spacefactor3000\relax}%
\providecommand \BibitemShut  [1]{\csname bibitem#1\endcsname}%
\let\auto@bib@innerbib\@empty
%</preamble>
\bibitem [{\citenamefont {Newman}(2010)}]{NewmanBook2010}%
  \BibitemOpen
  \bibfield  {author} {\bibinfo {author} {\bibfnamefont {M.}~\bibnamefont
  {Newman}},\ }\href@noop {} {\emph {\bibinfo {title} {Networks: An
  Introduction}}}\ (\bibinfo  {publisher} {Oxford University Press, Inc.},\
  \bibinfo {address} {New York, NY, USA},\ \bibinfo {year} {2010})\BibitemShut
  {NoStop}%
\bibitem [{\citenamefont {Albert}\ and\ \citenamefont
  {Barabasi}(2002)}]{Albert:2002p869}%
  \BibitemOpen
  \bibfield  {author} {\bibinfo {author} {\bibfnamefont {R.}~\bibnamefont
  {Albert}}\ and\ \bibinfo {author} {\bibfnamefont {A.}~\bibnamefont
  {Barabasi}},\ }\href {\doibase 10.1103/RevModPhys.74.47} {\bibfield
  {journal} {\bibinfo  {journal} {Rev. Mod. Phys.}\ }\textbf {\bibinfo {volume}
  {74}},\ \bibinfo {pages} {47} (\bibinfo {year} {2002})}\BibitemShut {NoStop}%
\bibitem [{\citenamefont {Newman}(2003)}]{Newman:2003p839}%
  \BibitemOpen
  \bibfield  {author} {\bibinfo {author} {\bibfnamefont {M.}~\bibnamefont
  {Newman}},\ }\href {\doibase 10.1137/S003614450342480} {\bibfield  {journal}
  {\bibinfo  {journal} {SIAM Review}\ }\textbf {\bibinfo {volume} {45}},\
  \bibinfo {pages} {167} (\bibinfo {year} {2003})}\BibitemShut {NoStop}%
\bibitem [{\citenamefont {Ugander}\ \emph {et~al.}()\citenamefont {Ugander},
  \citenamefont {Karrer}, \citenamefont {Backstrom},\ and\ \citenamefont
  {Marlow}}]{UganderFacebook}%
  \BibitemOpen
  \bibfield  {author} {\bibinfo {author} {\bibfnamefont {J.}~\bibnamefont
  {Ugander}}, \bibinfo {author} {\bibfnamefont {B.}~\bibnamefont {Karrer}},
  \bibinfo {author} {\bibfnamefont {L.}~\bibnamefont {Backstrom}}, \ and\
  \bibinfo {author} {\bibfnamefont {C.}~\bibnamefont {Marlow}},\ }\href@noop {}
  {}\bibinfo {note} {ArXiv:1111.4503}\BibitemShut {NoStop}%
\bibitem [{\citenamefont {Hill}\ and\ \citenamefont {Braha}(2010)}]{Hill2010}%
  \BibitemOpen
  \bibfield  {author} {\bibinfo {author} {\bibfnamefont {S.~A.}\ \bibnamefont
  {Hill}}\ and\ \bibinfo {author} {\bibfnamefont {D.}~\bibnamefont {Braha}},\
  }\href {\doibase 10.1103/PhysRevE.82.046105} {\bibfield  {journal} {\bibinfo
  {journal} {Phys. Rev. E}\ }\textbf {\bibinfo {volume} {82}},\ \bibinfo
  {pages} {046105} (\bibinfo {year} {2010})}\BibitemShut {NoStop}%
\bibitem [{\citenamefont {Barabasi}(2005)}]{Barabasi:2005aa}%
  \BibitemOpen
  \bibfield  {author} {\bibinfo {author} {\bibfnamefont {A.-L.}\ \bibnamefont
  {Barabasi}},\ }\href@noop {} {\bibfield  {journal} {\bibinfo  {journal}
  {Nature}\ }\textbf {\bibinfo {volume} {435}},\ \bibinfo {pages} {207}
  (\bibinfo {year} {2005})}\BibitemShut {NoStop}%
\bibitem [{\citenamefont {V\'azquez}\ \emph {et~al.}(2006)\citenamefont
  {V\'azquez}, \citenamefont {Oliveira}, \citenamefont {Dezs\"o}, \citenamefont
  {Goh}, \citenamefont {Kondor},\ and\ \citenamefont
  {Barab\'asi}}]{Alexei2006PRE}%
  \BibitemOpen
  \bibfield  {author} {\bibinfo {author} {\bibfnamefont {A.}~\bibnamefont
  {V\'azquez}}, \bibinfo {author} {\bibfnamefont {J.~G.}\ \bibnamefont
  {Oliveira}}, \bibinfo {author} {\bibfnamefont {Z.}~\bibnamefont {Dezs\"o}},
  \bibinfo {author} {\bibfnamefont {K.-I.}\ \bibnamefont {Goh}}, \bibinfo
  {author} {\bibfnamefont {I.}~\bibnamefont {Kondor}}, \ and\ \bibinfo {author}
  {\bibfnamefont {A.-L.}\ \bibnamefont {Barab\'asi}},\ }\href {\doibase
  10.1103/PhysRevE.73.036127} {\bibfield  {journal} {\bibinfo  {journal} {Phys.
  Rev. E}\ }\textbf {\bibinfo {volume} {73}},\ \bibinfo {pages} {036127}
  (\bibinfo {year} {2006})}\BibitemShut {NoStop}%
\bibitem [{\citenamefont {Barab\'{a}si}\ and\ \citenamefont
  {Albert}(1999)}]{Barabasi:1999p178}%
  \BibitemOpen
  \bibfield  {author} {\bibinfo {author} {\bibfnamefont {A.}~\bibnamefont
  {Barab\'{a}si}}\ and\ \bibinfo {author} {\bibfnamefont {R.}~\bibnamefont
  {Albert}},\ }\href {\doibase 10.1126/science.286.5439.509} {\bibfield
  {journal} {\bibinfo  {journal} {Science}\ }\textbf {\bibinfo {volume}
  {286}},\ \bibinfo {pages} {509} (\bibinfo {year} {1999})}\BibitemShut
  {NoStop}%
\bibitem [{\citenamefont {Caldarelli}\ \emph {et~al.}(2002)\citenamefont
  {Caldarelli}, \citenamefont {Capocci}, \citenamefont {De~Los~Rios},\ and\
  \citenamefont {Mu\~noz}}]{Caldarelli2002PRL}%
  \BibitemOpen
  \bibfield  {author} {\bibinfo {author} {\bibfnamefont {G.}~\bibnamefont
  {Caldarelli}}, \bibinfo {author} {\bibfnamefont {A.}~\bibnamefont {Capocci}},
  \bibinfo {author} {\bibfnamefont {P.}~\bibnamefont {De~Los~Rios}}, \ and\
  \bibinfo {author} {\bibfnamefont {M.~A.}\ \bibnamefont {Mu\~noz}},\ }\href
  {\doibase 10.1103/PhysRevLett.89.258702} {\bibfield  {journal} {\bibinfo
  {journal} {Phys. Rev. Lett.}\ }\textbf {\bibinfo {volume} {89}},\ \bibinfo
  {pages} {258702} (\bibinfo {year} {2002})}\BibitemShut {NoStop}%
\bibitem [{\citenamefont {Papadopoulos}\ \emph {et~al.}(2012)\citenamefont
  {Papadopoulos}, \citenamefont {Kitsak}, \citenamefont {Serrano},
  \citenamefont {Boguna},\ and\ \citenamefont
  {Krioukov}}]{Papadopoulos:2012aa}%
  \BibitemOpen
  \bibfield  {author} {\bibinfo {author} {\bibfnamefont {F.}~\bibnamefont
  {Papadopoulos}}, \bibinfo {author} {\bibfnamefont {M.}~\bibnamefont
  {Kitsak}}, \bibinfo {author} {\bibfnamefont {M.~A.}\ \bibnamefont {Serrano}},
  \bibinfo {author} {\bibfnamefont {M.}~\bibnamefont {Boguna}}, \ and\ \bibinfo
  {author} {\bibfnamefont {D.}~\bibnamefont {Krioukov}},\ }\href@noop {}
  {\bibfield  {journal} {\bibinfo  {journal} {Nature}\ }\textbf {\bibinfo
  {volume} {489}},\ \bibinfo {pages} {537} (\bibinfo {year}
  {2012})}\BibitemShut {NoStop}%
\bibitem [{\citenamefont {Muchnik}\ \emph {et~al.}(2013)\citenamefont
  {Muchnik}, \citenamefont {Pei}, \citenamefont {Parra}, \citenamefont {Reis},
  \citenamefont {Andrade~Jr}, \citenamefont {Havlin},\ and\ \citenamefont
  {Makse}}]{Muchnik2013}%
  \BibitemOpen
  \bibfield  {author} {\bibinfo {author} {\bibfnamefont {L.}~\bibnamefont
  {Muchnik}}, \bibinfo {author} {\bibfnamefont {S.}~\bibnamefont {Pei}},
  \bibinfo {author} {\bibfnamefont {L.~C.}\ \bibnamefont {Parra}}, \bibinfo
  {author} {\bibfnamefont {S.~D.~S.}\ \bibnamefont {Reis}}, \bibinfo {author}
  {\bibfnamefont {J.~S.}\ \bibnamefont {Andrade~Jr}}, \bibinfo {author}
  {\bibfnamefont {S.}~\bibnamefont {Havlin}}, \ and\ \bibinfo {author}
  {\bibfnamefont {H.~A.}\ \bibnamefont {Makse}},\ }\href@noop {} {\bibfield
  {journal} {\bibinfo  {journal} {Sci. Rep.}\ }\textbf {\bibinfo {volume}
  {3}},\ \bibinfo {pages} {1783} (\bibinfo {year} {2013})}\BibitemShut
  {NoStop}%
\bibitem [{\citenamefont {Vajna}\ \emph {et~al.}(2013)\citenamefont {Vajna},
  \citenamefont {T{\'o}th},\ and\ \citenamefont {Kert{\'e}sz}}]{Vajna2013}%
  \BibitemOpen
  \bibfield  {author} {\bibinfo {author} {\bibfnamefont {S.}~\bibnamefont
  {Vajna}}, \bibinfo {author} {\bibfnamefont {B.}~\bibnamefont {T{\'o}th}}, \
  and\ \bibinfo {author} {\bibfnamefont {J.}~\bibnamefont {Kert{\'e}sz}},\
  }\href@noop {} {\bibfield  {journal} {\bibinfo  {journal} {New J. Phys.}\
  }\textbf {\bibinfo {volume} {15}},\ \bibinfo {pages} {103023} (\bibinfo
  {year} {2013})}\BibitemShut {NoStop}%
\bibitem [{\citenamefont {Malmgren}\ \emph {et~al.}(2008)\citenamefont
  {Malmgren}, \citenamefont {Stouffer}, \citenamefont {Motter},\ and\
  \citenamefont {Amaral}}]{Malmgren25112008}%
  \BibitemOpen
  \bibfield  {author} {\bibinfo {author} {\bibfnamefont {R.~D.}\ \bibnamefont
  {Malmgren}}, \bibinfo {author} {\bibfnamefont {D.~B.}\ \bibnamefont
  {Stouffer}}, \bibinfo {author} {\bibfnamefont {A.~E.}\ \bibnamefont
  {Motter}}, \ and\ \bibinfo {author} {\bibfnamefont {L.~A.~N.}\ \bibnamefont
  {Amaral}},\ }\href {\doibase 10.1073/pnas.0800332105} {\bibfield  {journal}
  {\bibinfo  {journal} {Proc. Natl Proc. Acad. Sci. USA}\ }\textbf {\bibinfo
  {volume} {105}},\ \bibinfo {pages} {18153} (\bibinfo {year}
  {2008})}\BibitemShut {NoStop}%
\bibitem [{\citenamefont {{C. A. Hidalgo R.}}(2006)}]{HidalgoR2006877}%
  \BibitemOpen
  \bibfield  {author} {\bibinfo {author} {\bibnamefont {{C. A. Hidalgo R.}}},\
  }\href {\doibase http://dx.doi.org/10.1016/j.physa.2005.12.035} {\bibfield
  {journal} {\bibinfo  {journal} {Physica A}\ }\textbf {\bibinfo {volume}
  {369}},\ \bibinfo {pages} {877 } (\bibinfo {year} {2006})}\BibitemShut
  {NoStop}%
\bibitem [{\citenamefont {Sayama}\ \emph {et~al.}(2013)\citenamefont {Sayama},
  \citenamefont {Pestov}, \citenamefont {Schmidt}, \citenamefont {Bush},
  \citenamefont {Wong}, \citenamefont {Yamanoi},\ and\ \citenamefont
  {Gross}}]{Sayama20131645}%
  \BibitemOpen
  \bibfield  {author} {\bibinfo {author} {\bibfnamefont {H.}~\bibnamefont
  {Sayama}}, \bibinfo {author} {\bibfnamefont {I.}~\bibnamefont {Pestov}},
  \bibinfo {author} {\bibfnamefont {J.}~\bibnamefont {Schmidt}}, \bibinfo
  {author} {\bibfnamefont {B.~J.}\ \bibnamefont {Bush}}, \bibinfo {author}
  {\bibfnamefont {C.}~\bibnamefont {Wong}}, \bibinfo {author} {\bibfnamefont
  {J.}~\bibnamefont {Yamanoi}}, \ and\ \bibinfo {author} {\bibfnamefont
  {T.}~\bibnamefont {Gross}},\ }\href {\doibase
  http://dx.doi.org/10.1016/j.camwa.2012.12.005} {\bibfield  {journal}
  {\bibinfo  {journal} {Comput. Math. Appl.}\ }\textbf {\bibinfo {volume}
  {65}},\ \bibinfo {pages} {1645 } (\bibinfo {year} {2013})}\BibitemShut
  {NoStop}%
\bibitem [{\citenamefont {Gross}\ and\ \citenamefont
  {Blasius}(2008)}]{Gross:2008p274}%
  \BibitemOpen
  \bibfield  {author} {\bibinfo {author} {\bibfnamefont {T.}~\bibnamefont
  {Gross}}\ and\ \bibinfo {author} {\bibfnamefont {B.}~\bibnamefont
  {Blasius}},\ }\href {\doibase 10.1098/rsif.2007.1229} {\bibfield  {journal}
  {\bibinfo  {journal} {J. R. Soc. Interface}\ }\textbf {\bibinfo {volume}
  {5}},\ \bibinfo {pages} {259} (\bibinfo {year} {2008})}\BibitemShut {NoStop}%
\bibitem [{\citenamefont {Holme}\ and\ \citenamefont
  {Saram{\"a}ki}(2012)}]{Holme201297}%
  \BibitemOpen
  \bibfield  {author} {\bibinfo {author} {\bibfnamefont {P.}~\bibnamefont
  {Holme}}\ and\ \bibinfo {author} {\bibfnamefont {J.}~\bibnamefont
  {Saram{\"a}ki}},\ }\href {\doibase
  http://dx.doi.org/10.1016/j.physrep.2012.03.001} {\bibfield  {journal}
  {\bibinfo  {journal} {Phys. Rep.}\ }\textbf {\bibinfo {volume} {519}},\
  \bibinfo {pages} {97 } (\bibinfo {year} {2012})}\BibitemShut {NoStop}%
\bibitem [{\citenamefont {Holme}(2015)}]{Holme2015}%
  \BibitemOpen
  \bibfield  {author} {\bibinfo {author} {\bibfnamefont {P.}~\bibnamefont
  {Holme}},\ }\href@noop {} {\bibfield  {journal} {\bibinfo  {journal} {Eur.
  Phys. J. B}\ }\textbf {\bibinfo {volume} {88}},\ \bibinfo {pages} {234}
  (\bibinfo {year} {2015})}\BibitemShut {NoStop}%
\bibitem [{\citenamefont {Zhao}\ \emph {et~al.}(2011)\citenamefont {Zhao},
  \citenamefont {Stehl\'e}, \citenamefont {Bianconi},\ and\ \citenamefont
  {Barrat}}]{Zha02011}%
  \BibitemOpen
  \bibfield  {author} {\bibinfo {author} {\bibfnamefont {K.}~\bibnamefont
  {Zhao}}, \bibinfo {author} {\bibfnamefont {J.}~\bibnamefont {Stehl\'e}},
  \bibinfo {author} {\bibfnamefont {G.}~\bibnamefont {Bianconi}}, \ and\
  \bibinfo {author} {\bibfnamefont {A.}~\bibnamefont {Barrat}},\ }\href
  {\doibase 10.1103/PhysRevE.83.056109} {\bibfield  {journal} {\bibinfo
  {journal} {Phys. Rev. E}\ }\textbf {\bibinfo {volume} {83}},\ \bibinfo
  {pages} {056109} (\bibinfo {year} {2011})}\BibitemShut {NoStop}%
\bibitem [{\citenamefont {Starnini}\ \emph {et~al.}(2013)\citenamefont
  {Starnini}, \citenamefont {Baronchelli},\ and\ \citenamefont
  {Pastor-Satorras}}]{Starnini2013}%
  \BibitemOpen
  \bibfield  {author} {\bibinfo {author} {\bibfnamefont {M.}~\bibnamefont
  {Starnini}}, \bibinfo {author} {\bibfnamefont {A.}~\bibnamefont
  {Baronchelli}}, \ and\ \bibinfo {author} {\bibfnamefont {R.}~\bibnamefont
  {Pastor-Satorras}},\ }\href {\doibase 10.1103/PhysRevLett.110.168701}
  {\bibfield  {journal} {\bibinfo  {journal} {Phys. Rev. Lett.}\ }\textbf
  {\bibinfo {volume} {110}},\ \bibinfo {pages} {168701} (\bibinfo {year}
  {2013})}\BibitemShut {NoStop}%
\bibitem [{\citenamefont {Rocha}\ \emph {et~al.}(2011)\citenamefont {Rocha},
  \citenamefont {Liljeros},\ and\ \citenamefont {Holme}}]{Rocha:2011bq}%
  \BibitemOpen
  \bibfield  {author} {\bibinfo {author} {\bibfnamefont {L.~E.~C.}\
  \bibnamefont {Rocha}}, \bibinfo {author} {\bibfnamefont {F.}~\bibnamefont
  {Liljeros}}, \ and\ \bibinfo {author} {\bibfnamefont {P.}~\bibnamefont
  {Holme}},\ }\href@noop {} {\bibfield  {journal} {\bibinfo  {journal} {PLoS
  Comput. Biol.}\ }\textbf {\bibinfo {volume} {7}},\ \bibinfo {pages}
  {e1001109} (\bibinfo {year} {2011})}\BibitemShut {NoStop}%
\bibitem [{\citenamefont {Perra}\ \emph {et~al.}(2012)\citenamefont {Perra},
  \citenamefont {Gon{\c c}alves}, \citenamefont {Pastor-Satorras},\ and\
  \citenamefont {Vespignani}}]{Perra:2012fi}%
  \BibitemOpen
  \bibfield  {author} {\bibinfo {author} {\bibfnamefont {N.}~\bibnamefont
  {Perra}}, \bibinfo {author} {\bibfnamefont {B.}~\bibnamefont {Gon{\c
  c}alves}}, \bibinfo {author} {\bibfnamefont {R.}~\bibnamefont
  {Pastor-Satorras}}, \ and\ \bibinfo {author} {\bibfnamefont {A.}~\bibnamefont
  {Vespignani}},\ }\href@noop {} {\bibfield  {journal} {\bibinfo  {journal}
  {Sci. Rep.}\ }\textbf {\bibinfo {volume} {2}},\ \bibinfo {pages} {469}
  (\bibinfo {year} {2012})}\BibitemShut {NoStop}%
\bibitem [{\citenamefont {Karsai}\ \emph {et~al.}(2014)\citenamefont {Karsai},
  \citenamefont {Perra},\ and\ \citenamefont {Vespignani}}]{Karsai:2014aa}%
  \BibitemOpen
  \bibfield  {author} {\bibinfo {author} {\bibfnamefont {M.}~\bibnamefont
  {Karsai}}, \bibinfo {author} {\bibfnamefont {N.}~\bibnamefont {Perra}}, \
  and\ \bibinfo {author} {\bibfnamefont {A.}~\bibnamefont {Vespignani}},\
  }\href {\doibase 10.1038/srep04001} {\bibfield  {journal} {\bibinfo
  {journal} {Sci. Rep.}\ }\textbf {\bibinfo {volume} {4}},\ \bibinfo {pages}
  {4001} (\bibinfo {year} {2014})}\BibitemShut {NoStop}%
\bibitem [{\citenamefont {Vestergaard}\ \emph {et~al.}(2014)\citenamefont
  {Vestergaard}, \citenamefont {G\'enois},\ and\ \citenamefont
  {Barrat}}]{Vestergaard2014}%
  \BibitemOpen
  \bibfield  {author} {\bibinfo {author} {\bibfnamefont {C.~L.}\ \bibnamefont
  {Vestergaard}}, \bibinfo {author} {\bibfnamefont {M.}~\bibnamefont
  {G\'enois}}, \ and\ \bibinfo {author} {\bibfnamefont {A.}~\bibnamefont
  {Barrat}},\ }\href {\doibase 10.1103/PhysRevE.90.042805} {\bibfield
  {journal} {\bibinfo  {journal} {Phys. Rev. E}\ }\textbf {\bibinfo {volume}
  {90}},\ \bibinfo {pages} {042805} (\bibinfo {year} {2014})}\BibitemShut
  {NoStop}%
\bibitem [{\citenamefont {Rocha}\ \emph {et~al.}(2010)\citenamefont {Rocha},
  \citenamefont {Liljeros},\ and\ \citenamefont {Holme}}]{RochaPNAS2010}%
  \BibitemOpen
  \bibfield  {author} {\bibinfo {author} {\bibfnamefont {L.~E.~C.}\
  \bibnamefont {Rocha}}, \bibinfo {author} {\bibfnamefont {F.}~\bibnamefont
  {Liljeros}}, \ and\ \bibinfo {author} {\bibfnamefont {P.}~\bibnamefont
  {Holme}},\ }\href {\doibase 10.1073/pnas.0914080107} {\bibfield  {journal}
  {\bibinfo  {journal} {Proc. Natl Proc. Acad. Sci. USA}\ }\textbf {\bibinfo
  {volume} {107}},\ \bibinfo {pages} {5706} (\bibinfo {year}
  {2010})}\BibitemShut {NoStop}%
\bibitem [{\citenamefont {Holme}(2003)}]{Holme2003EPL}%
  \BibitemOpen
  \bibfield  {author} {\bibinfo {author} {\bibfnamefont {P.}~\bibnamefont
  {Holme}},\ }\href@noop {} {\bibfield  {journal} {\bibinfo  {journal}
  {Europhys. Lett.}\ }\textbf {\bibinfo {volume} {64}},\ \bibinfo {pages} {427}
  (\bibinfo {year} {2003})}\BibitemShut {NoStop}%
\bibitem [{Note1()}]{Note1}%
  \BibitemOpen
  \bibinfo {note} {See Supplemental Material at [URL will be inserted by
  publisher] for the movie}\BibitemShut {NoStop}%
\bibitem [{\citenamefont {Wilf}(2006)}]{Generatingfunctionology}%
  \BibitemOpen
  \bibfield  {author} {\bibinfo {author} {\bibfnamefont {H.~S.}\ \bibnamefont
  {Wilf}},\ }\href@noop {} {\emph {\bibinfo {title}
  {Generatingfunctionology}}}\ (\bibinfo  {publisher} {A. K. Peters, Ltd.},\
  \bibinfo {address} {Natick, MA, USA},\ \bibinfo {year} {2006})\BibitemShut
  {NoStop}%
\bibitem [{Note2()}]{Note2}%
  \BibitemOpen
  \bibinfo {note} {See the supplementary material for details on the generating
  function}\BibitemShut {NoStop}%
\bibitem [{\citenamefont {Boccaletti}\ \emph {et~al.}(2006)\citenamefont
  {Boccaletti}, \citenamefont {Latora}, \citenamefont {Moreno}, \citenamefont
  {Chavez},\ and\ \citenamefont {Hwang}}]{Boccaletti2006175}%
  \BibitemOpen
  \bibfield  {author} {\bibinfo {author} {\bibfnamefont {S.}~\bibnamefont
  {Boccaletti}}, \bibinfo {author} {\bibfnamefont {V.}~\bibnamefont {Latora}},
  \bibinfo {author} {\bibfnamefont {Y.}~\bibnamefont {Moreno}}, \bibinfo
  {author} {\bibfnamefont {M.}~\bibnamefont {Chavez}}, \ and\ \bibinfo {author}
  {\bibfnamefont {D.-U.}\ \bibnamefont {Hwang}},\ }\href {\doibase
  http://dx.doi.org/10.1016/j.physrep.2005.10.009} {\bibfield  {journal}
  {\bibinfo  {journal} {Phys. Rep.}\ }\textbf {\bibinfo {volume} {424}},\
  \bibinfo {pages} {175 } (\bibinfo {year} {2006})}\BibitemShut {NoStop}%
\bibitem [{\citenamefont {Clauset}\ \emph {et~al.}(2004)\citenamefont
  {Clauset}, \citenamefont {Newman},\ and\ \citenamefont
  {Moore}}]{Clauset:2004dz}%
  \BibitemOpen
  \bibfield  {author} {\bibinfo {author} {\bibfnamefont {A.}~\bibnamefont
  {Clauset}}, \bibinfo {author} {\bibfnamefont {M.E.J}~\bibnamefont {Newman}}, \
  and\ \bibinfo {author} {\bibfnamefont {C.}~\bibnamefont {Moore}},\
  }\href@noop {} {\bibfield  {journal} {\bibinfo  {journal} {Phys. Rev. E}\
  }\textbf {\bibinfo {volume} {70}},\ \bibinfo {pages} {066111} (\bibinfo
  {year} {2004})}\BibitemShut {NoStop}%
\bibitem [{\citenamefont {Newman}(2006)}]{Newman:2006iq}%
  \BibitemOpen
  \bibfield  {author} {\bibinfo {author} {\bibfnamefont {M.}~\bibnamefont
  {Newman}},\ }\href@noop {} {\bibfield  {journal} {\bibinfo  {journal} {Proc.
  Natl Proc. Acad. Sci. USA}\ }\textbf {\bibinfo {volume} {103}},\ \bibinfo
  {pages} {8577} (\bibinfo {year} {2006})}\BibitemShut {NoStop}%
\bibitem [{\citenamefont {Fortunato}(2010)}]{Fortunato:2010iw}%
  \BibitemOpen
  \bibfield  {author} {\bibinfo {author} {\bibfnamefont {S.}~\bibnamefont
  {Fortunato}},\ }\href@noop {} {\bibfield  {journal} {\bibinfo  {journal}
  {Phys. Rep.}\ }\textbf {\bibinfo {volume} {486}},\ \bibinfo {pages} {75}
  (\bibinfo {year} {2010})}\BibitemShut {NoStop}%
\end{thebibliography}
\end{document}